 \documentstyle[12pt]{article}

 \newcommand{\pa}{\partial}
 \newcommand{\Tt}{\tilde{T}}
 \newcommand{\Ut}{\tilde{U}}
 \newcommand{\Vt}{\tilde{V}}
 \newcommand{\ct}{\tilde{c}}
 \newcommand{\ep}{\epsilon}
 
 \newcommand{\strutje}{\rule[-1.5mm]{0mm}{5mm}}
 \newcommand{\str}{\rule[-2.5mm]{0mm}{7mm}}

 \newcommand{\extraspace}{\addtolength{\abovedisplayskip}{2mm}
			 \addtolength{\belowdisplayskip}{2mm}
			 \addtolength{\abovedisplayshortskip}{2mm}
			 \addtolength{\belowdisplayshortskip}{2mm}}
 \newcommand{\be}{\begin{equation}\extraspace}
 
 \newcommand{\ee}{\end{equation}}
 
 \newcommand{\bea}{\begin{eqnarray}\extraspace}
 \newcommand{\beastar}{\begin{eqnarray*}\extraspace}
 \newcommand{\eea}{\end{eqnarray}}
 \newcommand{\eeastar}{\end{eqnarray*}}
 \newcommand{\nonu}{\nonumber \\[2mm]}
 \newcommand{\np}{Nucl.Phys.\ }
 
 \newcommand{\cmp}{Comm.Math.Phys.\ }
 \newcommand{\pl}{Phys.Lett.\ }

\begin{document}

 \begin{flushright}
				      ITP.SB-92-45 \\ Sept.'92
 \end{flushright}

 \vspace{5mm}

 \begin{center}
 {\LARGE Explicit Construction  of Spin 4 Casimir Operator in the Coset
 Model $ \hat{SO} (5)_{1} \times \hat{SO} (5)_{m} / \hat{SO} (5)_{1+m} $}\\

 \vspace{1cm}

 {\large Changhyun Ahn\footnote{email: ahn@max.physics.sunysb.edu}
 \footnote{ Researcher, Natural Science Institute, Yonsei University,
 Seoul, 120-749, Korea}} \\

 \vspace{7mm}
 \str {\em Institute for Theoretical Physics} \\
 \str {\em State University of New York at Stony Brook} \\
 \str {\em Stony Brook, NY 11794-3840} \\
 \vspace{2cm}
 \end{center}

 We generalize the Goddard-Kent-Olive (GKO) coset construction to the
 dimension 5/2 operator for $ \hat{so} (5) $ and compute the fourth order
 Casimir invariant in the coset
 model $\hat{SO} (5)_{1} \times \hat{SO} (5)_{m} / \hat{SO} (5)_{1+m} $
 with the generic unitary minimal  $ c < 5/2 $ series
 that can be viewed as perturbations of the $ m \rightarrow \infty $ limit,
 which has been
 investigated previously in the realization of $ c= 5/2 $ free fermion model.

 \vspace{3mm}

 \baselineskip=18pt

 \newpage


 \baselineskip=18pt

 \indent

 Extensions of the conformal symmetry have played an important role
 in making a systematic study of two dimensional conformal quantum
 field theories. It is of great consequence to realize that in a given
 model of conformal field theory the true symmetry algebra is bigger
 than the conformal algebra alone. The representations of such model can
 be enhanced by studying the extra symmetry. Very recently, non-linear
 higher spin extensions of the Virasoro algebra, so-called $W$-algebras,
 have been reviewed in \cite{bs}.

 After 2 scalar free field realizations \cite{fz} of Zamolodchikov's
 $ W_{3} $ algebra \cite{zamo}, Fateev and Lukyanov \cite{fl2} have extended
 it to
 the infinite dimensional
 associative algebras $ WA_{n}, WD_{n} \;\mbox{and} \;WB_{n},$ based on the
 finite Lie algebras $ A_{n}, D_{n} \;\mbox{and} \;B_{n}, $ respectively. In
 particular, the spins of the fields of the $ WB_{n}$ algebra are  not
 related to the exponents of $ B_{n}= so(2n+1) $, but instead to those
 of the Lie
 superalgebra $ B(0,n)=osp(1,2n)$  $ ; 1, 3,
 \cdots, 2n-1 $ and $ n-1/2 $ ( the exponents plus 1 ) \cite{watts}.
 For $n=1$, the
 $WB_{1}$
 algebra corresponds to the $ N=1$ superconformal algebra.

 Extended symmetries are easy to deal with in conformal field theories based on
the
 Goddard-Kent-Olive (GKO) coset constuction \cite{gko}. In this approach,
 the unitary representations of Casimir algebra ( $ c < l, $ where $l$ is the
 rank of the algebra ) can be obtained from the cosets for the algebras $ADE$.
 For the non simply laced algebra, $B_{n}$, it was further discussed in \cite
 {watts}. The discrete series of $c$ values \cite{fl2} from free field
 construction coincide with values obtained from coset models associated with
 $\hat{B_{n}}=\hat{so} (2n+1)$ algebra.

 The existence of $ WB_{2}$ algebra which is associative for all values of
 $c$,
 explicitly using the perturbative conformal
 bootstrap, has been shown in \cite{fst}. We were able to reproduce their
findings by exploiting the graded Jacobi identity for the Laurent expansion
modes
 of generating currents and
 illustrated a realization of $ c= 5/2 $ free fermion model from the
 basic fermion fields and finally confirmed that the $bosonic$ currents in the
 $ WB_{2} $ algebra are the Casimirs of $\hat{so} (5) $ \cite{ahn}.

 In this paper, we want to generalize the algebraic structure of \cite{ahn}
 and construct $ WB_{2} $ currents in the coset model of the generic
 unitary minimal discrete $ c < 5/2 $ series and explain briefly how to
 count the $so(5)$ singlets in the $ c=5/2$ free fermion model by analyzing
 a generating function.

 We take a look at the coset model $
 \hat{SO} (5)_{1} \times \hat{SO} (5)_{m} / \hat{SO} (5)_{1+m} $
 which can be regarded as perturbations of the $ m \rightarrow \infty $
 model that was considered in \cite{ahn}.
 Denoting the generators of the algebra $ g=\hat{so} (5) \oplus \hat{
 so} (5) $ as $E_{(1)}^{ab}(z)$ and $E_{(2)}^{ab}(z)$, of level
 $1$ and $m$, respectively, and those of the diagonal subalgebra $g^{\prime}
 =\hat{so} (5)$, which has level $m^{\prime}=1+m$, as
 $ {E^{\prime}}^{ab} (z),$ we have the relation,
 \be
 {E^{\prime}}^{ab}(z)=E_{(1)}^{ab}(z)+E_{(2)}^{ab}(z).
 \ee
 The indices $a$ and $b$ take values in the adjoint representation of
 $so(5)$ and $ a,b=1, 2, \cdots, 5.$
 $\hat{so} (5)$ algebra has 10 fields ( $ E^{ab} (z)=-E^{ba} (z) $ ).
 The coset Virasoro generator $\tilde{T} (z) $ is given by, as usual,
 \bea
  \tilde {T} (z) &=& T_{(1)}(z)+T_{(2)}(z)-T^{\prime}(z)\nonu
		 &=&-\frac{1}{16} E_{(1)}^{ab} E_{(1)}^{ab}(z)-\frac{1}{4(m+3)
 } E_{(2)}^{ab} E_{(2)}^{ab}(z)+\frac{1}{4(m+4)} {E^{\prime}}^{ab}
 {E^{\prime}}^{ab}(z),
 \eea
 which commutes with ${E^{\prime}}^{ab}(z)$.
 The coset central charge of the unitary minimal models for $WB_{2}$
 is
 \be
 \tilde{c}=c(WB_{2})=\frac{5}{2}+\frac{10m}{m+3}-\frac{10(m+1)}{m+4}
 =\frac{5}{2} [1-\frac{12}{(m+3)(m+4)}],
 \ee
 where $  m=1,2, \cdots .$

 Our next step is to extend the GKO coset construction to the dimension 5/2
 operator associated with $\hat{so} (5)$ and we follow the analysis in
 \cite{bais}. In order to write down the coset analogue $\Ut (z)$ of dimension
5/2,
 we allow a general
 linear combination of the terms,
 \be
  \ep^{abcde} \psi^{a}_{(1)} E^{bc}_{(1)} E^{de}_
 {(1)},\;\;
 \ep^{abcde} \psi^{a}_{(1)} E^{bc}_{(1)} E^{de}_{(2)} \;\;\mbox{and} \;\;
 \ep^{abcde} \psi^{a}_{(1)} E^{bc}_{(2)} E^{de}_{(2)},
 \ee
 and proceed by imposing that $\Ut(z) $
 transforms under $\tilde {T}(z) $ as dimension $5/2$ primary field. Then
 we have found that it uniquely fixes $\Ut(z) $ up to a normalization
 factor $A (1, m)$.
 \be
 \Ut(z) = A (1, m) \ep^{abcde} [ \psi^{a}_{(1)} E^{bc}_{(1)} E^{de}_{(1)}-
 \frac{10}{m} \psi^{a}_{(1)} E^{bc}_{(1)}
 E^{de}_{(2)} +\frac{15}{m(m+2)} \psi^{a}_{(1)} E^{bc}_{(2)} E^{de}_{(2)} ]
 (z),
 \ee
 which are singlets under the underlying $ {so} (5) $ subalgebra of
 $g^{\prime}=\hat{so} (5) $. One can show that $\Ut (z) $ has an element of
 the coset zero contraction with $ {E^{\prime}}^{ab} (z).$
 The normalization factor $A(1, m)$ can be fixed as
 \be
 A(1, m)=\frac{m}{120}\sqrt{\frac{(m+2)}{(m+3)(m+4)(m+5)}}\;\;,
 \ee
 by investigating $ 1/(z-w)^5$ term of $\Ut(z) \Ut(w)$. The expression
 for $\Ut (z) $ has been recognized also in \cite{watts} with different
 normalizations.
 One can easily see that $\Ut (z)$ reduces to $U(z)=\frac{1}{120} \ep^{abcde}
 \psi_{(1)}^{a} E_{(1)}^{bc} E_{(1)}^{de} $ \cite{ahn}
 as $m \rightarrow \infty $.
 After a tedious work, using repeatedly the Wick's theorem for
 the operator product
 expansion \cite{bais}, one finds
 \bea
 &&\Ut(z) \Ut(w) = \frac{1}{(z-w)^{5}} \frac{m(m+7)}{(m+3)(m+4)}\nonu
 &&+\frac{1}{(z-w)^{3}} \frac{1}{(m+4)}[-\frac{m}{8}
 E_{(1)}^{ab} E_{(1)}^{ab} + E_{(1)}^{ab} E_{(2)}^{ab}-\frac{1}{2(m+3)}
 E_{(2)}^{ab} E_{(2)}^{ab}] (w)\nonu
 && + \frac{1}{(z-w)^{2}}\frac{1}{(m+4)}[-\frac{m}{8} E_{(1)}^{ab}
 \pa E_{(1)}^{ab}+\frac{1}{2} \pa ( E_{(1)}^{ab} E_{(2)}^{ab} )-\frac{1}{
 2(m+3)} E_{(2)}^{ab} \pa E_{(2)}^{ab}] (w)\nonu
 &&+\frac{1}{(z-w)} \frac{m^2(m+2)}{14400(m+3)(m+4)(m+5)}
 [\frac{7200(m+2)}{m} \psi_{(1)}^{a} \pa \psi_{(1)}^{a} \psi_{(1)}^{b} \pa
  \psi_{(1)}^{b}\nonu
 &&-\frac{2400(m-1)(m+3)}{m(m+2)} \psi_{(1)}^{a} \pa^{3} \psi
 _{(1)}^{a}-\frac{100}{m} \ep^{abcde} \ep^{afghi} ( E_{(1)}^{bc} E_{(
 1)}^{de} )( E_{(1)}^{fg} E_{(2)}^{hi})\nonu
 &&-\frac{400(m+7)}{m^2} \ep^{abcde}
 \ep^{abfgh} \psi_{(1)}^{f} \pa ( \psi_{(1)}^{c} E_{(1)}^{de} ) E_{(2)}^{gh}
  -\frac{3600(m^2+5m+12)}{m^2(m+2)} \pa ^{2} E_{(1)}^{ab} E_{
 (2)}^{ab}\nonu
 &&+ \frac{3600(m+3)}{m^2} E_{(1)}^{ab} \pa^{2} E_{(2)}^{ab}
 +\frac{50(5m+16)}{m^2(m+2)} \ep^{abcde} \ep^{afghi} ( E_{(1)}
 ^{bc} E_{(1)}^{de} ) (E_{(2)}^{fg} E_{(2)}^{hi})\nonu
 &&+\frac{300}{m^2} \ep^{abcde}
 \ep^{afghi} (E_{(1)}^{bc} E_{(2)}^{de})(E_{(1)}^{fg} E_{(2)}^{hi})\nonu
 && +\frac{600(m+5)}{m^2(m+2)} \ep^{abcde} \ep^{abfgh} \psi_{(1)}
 ^{f} ( \pa (\psi_{(1)}^{c} E_{(2)}^{de} ) E_{(2)}^{gh})\nonu
 &&-\frac{900(m+3)}{m^2(
 m+2)^2} \ep^{abcde} \ep^{afghi} (E_{(1)}^{bc} E_{(2)}^{de})(E_{(2)}^{fg} E_{(
 2)}^{hi})\nonu
 &&+\frac{225}{m^2(m+2)^2} \ep^{abcde} \ep^{afghi} (E_{(2)}^{bc}
  E_{(2)}^{de})( E_{(2)}^{fg} E_{(2)}^{hi} )] (w)+ \cdots.
 \eea
 The parentheses denote the 'normal ordered product' between operators
 at coincident
 points. It can be defined as a contour integral for the operators $A$ and $B$.
 \bea
 (AB)(z)=\frac{1}{2 \pi i} \oint_{z} \; \frac{dw}{w-z} A(w) B(z).
 \eea

How can we extract the dimension 4 coset field  $ \Vt(z) $ from the singular
 part
 of $1/(z-w)$ in the above operator product expansion ? The convenient way to
do
 it is to make all the composite operators be 'fully normal ordered products'
 \cite{bais}.
 Then we arrive at the following results, using extensively the rearrangement
 lemmas \cite{bais},
 \bea
 & & \Ut(z) \Ut(w)  =  \frac{1}{(z-w)^{5}} \frac{2}{5} \ct +\frac{1}
 {(z-w)^{3}} 2\Tt(w)+\frac{1}{(z-w)^{2}} \pa \Tt(w) \nonu
 & & +\frac{1}{(z-w)} [\frac{3}{10} \pa ^{2} \Tt(w) +\frac{27}
 {(5\ct+22)} \tilde{\Lambda}(w)+\sqrt{\frac{6(14\ct+13)}{(5\ct+22)}}\; \Vt(w)]
 +\cdots,
 \eea
 and
 \bea
 &&\Vt(z) =\frac{1}{4(m+4)(m+5)\sqrt{3(2m+1)(2m+13)(23m^2+161m+176)}}\nonu
 &&\times [a \pa E_{(1)}
	^{ab} \pa E_{(1)}^{ab} +b E_{(1)}^{ab} \pa^{2} E_{(1)}^{ab}
	+t E_{(1)}^{ab} E_{(1)}^{ab} E_{(1)}^{cd} E_{(1)}^{cd}
	\nonu
 && +d \pa E_{(2)}^{ab} \pa E_{(2)}^{ab}+e E_{(2)}^{ab} \pa^{2} E_{(2)}^{
	 ab}+f \pa^{2} E_{(1)}^{ab} E_{(2)}^{ab}\nonu
 &&+g \pa E_{(1)}^{ab} \pa E_{(2)}^{ab}+h E_{(1)}^{ab} \pa^{2} E_{(2)}^
	{ab}+i E_{(1)}^{ab} E_{(1)}^{ab} E_{(1)}^{cd} E_{(2)}^{cd}\nonu
 &&+j E_{(1)}^{ab} \pa E_{(1)}^{ac} E_{(2)}^{bc}+k E_{(1)}^{ab} E_{(1)
	}^{ab} E_{(2)}^{cd} E_{(2)}^{cd}+l E_{(1)}^{ab} E_{(1)}^{cd} E_{(2)
	}^{ab} E_{(2)}^{cd}\nonu
 &&+s E_{(1)}^{ab} E_{(2)}^{ac} \pa E_{(2)}^{bc}+n E_{(1)}^{ab} E_{(2)}^
	{ab} E_{(2)}^{cd} E_{(2)}^{cd}+o E_{(2)}^{ab} E_{(2)}^{ab} E_{(2)}^
	{cd} E_{(2)}^{cd}\nonu
 &&+p E_{(1)}^{ab} E_{(1)}^{ac} E_{(2)}^{cd} E_{(2)}^{bd}+q E_{(1)}^
	{ab} E_{(2)}^{cd} E_{(2)}^{ac} E_{(2)}^{bd}+r E_{(2)}^{ab} E_{(2)}^
	{cd} E_{(2)}^{ac} E_{(2)}^{bd}](z),
 \eea
 where
 \bea
 & &a=\frac{3}{4} m(m+2)(2m+1)(3m+17),\; b=-\frac{1}{2}m(m+2)(2m+1)(3m+17)
 \nonu
 & &t=-\frac{1}{16} m(m+2)(2m+1)(2m+11)\nonu
 & & d=\frac{3(2m^{4}+28m^{3}+33m^{2}
 -455m-688)}{8(m+2)(m+3)}\nonu
 & &e=-\frac{(2m^{4}+28m^{3}+33m^{2}-455m-688)}{4(m+2)(m+3)}\nonu
 & & f=\frac{1}{
 4} (m+2)(2m+1)(45m+259),\;
 g=-\frac{3}{2} (m^{3}+58m^{2}+349m+312)\nonu
 & & h=\frac{(2m^{4}+43m^{2}-93m^{2}
 -2084m-2836)}{4(m+2)}\nonu
 & &i=(m+2)(2m+1)(2m+11),\; j=(m+2)(2m+1)(7m+41)\nonu
 & &k=\frac{3}{8} (10m^{2}+73m+88),\; l=\frac{1}{2} (2m+1)(7m+41)\nonu
 & &n=-\frac{(7m^{2}+49m+43)}{(m+2)},\; o=\frac{(7m^{2}+49m+43)}{4(m+2)(m+3)}
 \nonu
 & &p=-(23m^{2}+161m+176),\; q=\frac{(23m^{2}+161m+176)}{(m+2)}\nonu
 & &r=-\frac{(23m^{2}+161m+176)}{4(m+2)(m+3)},\; s=\frac{(5m^{3}-42m^{2}-543m
 -716)}{(m+2)}.
 \eea
 The result for $V(z)$ in \cite{ahn} are recovered in the limit $ m \rightarrow
 \infty $ :
 \bea
 \Vt (z) \;\; \rightarrow \;\; & &  \frac{1}{8\sqrt{69}} [ \frac{9}{2}
 \pa E_{(1)}^
 {ab} \pa E_{(1)}^{ab} -3 E_{(1)}^{ab} \pa^2 E_{(1)}^{ab} -\frac{1}{4} E_{(1)}
 ^{ab} E_{(1)}^{ab} E_{(1)}^{cd} E_{(1)}^{cd} ] (z)\nonu
			       & & = -\frac{1}{40\sqrt{69}} E_{(1)}^{ab} E_{(1)}^
 {cd} E_{(1)}^{ac} E_{(1)}^{bd} (z)= V(z).
 \eea
 All the operators are replaced by their coset analogues comparing with
the results
 of \cite{ahn} and the central charge
 has the value of the eq. (3). The one thing which we would like to stress is
 the fact that
 the 9 independent fields containing the derivatives in the above expression of
 $\Vt (z)$ actually can be reexpressed in terms of the products of 4
$E^{ab}$'s.
 As an example,
 \bea
 E_{(1)}^{ab} E_{(2)}^{ac} \pa E_{(2)}^{bc} (z) & = & -\frac{1}{3} [
E_{(1)}^{ab}
 E_{(2)}^{ac} E_{(2)}^{bd} E_{(2)}^{cd}- E_{(1)}^{ab} E_{(2)}^{cd} E_{(2)}^{
 ac} E_{(2)}^{bd} ](z)\nonu
					    &   & -\frac{1}{4} [ E_{(1)}^{ab}
 E_{(2)}^{cd} E_{(2)}^{ab} E_{(2)}^{cd}- E_{(1)}^{ab} E_{(2)}^{ab} E_{(2)}^{cd}
 E_{(2)}^{cd} ](z).
 \eea
 Finally, $\Vt(z)$ can be written as
 \bea
 & & [C_{1}(m) \delta^{ea} \delta^{fb} \delta^{gc} \delta^{hd}+C_{2}(m) \delta
^{ea} \delta^{fc} \delta^{gb} \delta^{hd}] E^{ab} E^{cd} E^{ef}
E^{gh}(z)\nonu
 & & +[C_{3}(m) \delta^{ea} \delta^{fb} \delta^{gc} \delta^{hd}+C_{4}(m) \delta
^{ea} \delta^{fc} \delta^{gb} \delta^{hd} ]E^{ab} E^{ef} E^{cd} E^{gh}(z)\nonu
 & & +C_{5}(m) \delta^{ea} \delta^{fc} \delta^{gb} \delta^{hd} E^{ab} E^{ef}
E^{gh} E^{cd}(z),
 \eea
 where $E^{ab}(z)$ is $E_{(1)}^{ab}(z)$ or $E_{(2)}^{ab}(z)$ and $C$'s are
 some functions of $m$ and $\delta^{ab}$ is an invariant tensor of $so(5)$.
 Therefore $\Vt (z)$ is really the fourth order Casimir operator for $\hat{so}
 (5)$.
 On the other hand, the field contents of $\pa^{2} \Tt (z) \;\;\mbox{and}
\;\;\tilde{\Lambda} (z)=\Tt^{2}(z)-\frac{3}{10} \pa^{2} \Tt(z)
 $ are the
 same as those of $\Vt (z) $ except that they do $not$
 have the following terms
 \be
  E^{ab}_{(1)}E^{ac}_{(1)} E^{cd}_{(2)} E^{bd}_{(2)},\;\; E^{ab}_{(1)}
E^{cd}_{(2)}
 E^{ac}_{(2)}E^{bd}_{(2)}\;\;
 \mbox {and} \;\;E^{ab}_{(2)} E^{cd}_{(2)} E^{ac}_{(2)} E^{bd}_{(2)}.
 \ee

 In order to check that $\Vt (z)$ is a dimension 4 primary field with
 respect to $\Tt(z)$, we should compute by explicit calculations the operator
 product expansions of $\Tt(z)$ with 18 fields eq. (10). As a check on this,
 the fact that $\Tt(z)$ commutes with ${E^{\prime}}^{ab}(z)$ has been used
 repeatedly. We list the operator product expansions $\Tt(z)$ with the
fields of (15).
 \bea
 &&\Tt(z) E_{(1)}^{ab} E_{(1)}^{ac} E_{(2)}^{cd} E_{(2)}^{bd}(w)  =  -\frac{1}{
 (z-w)^6} \frac{180m}{(m+4)}\nonu
 &&+\frac{1}{
 (z-w)^{4}} \frac{1}{(m+4)} [-4m E_{(2)}^{ab} E_{(2)}^{ab}-4m E_{(1)}^{ab}
E_{(1)}^{ab} +(19m+12) E_{(1)}^{ab} E_{(2)}^{ab} ](w)\nonu
 &&+\frac{
 1}{(z-w)^3} \frac{9}{2(m+4)} [-(m+3) E_{(1)}^{ab} \pa E_{(2)}^{ab}-
 4\pa E_{(1)}^{ab} E_{(2)}^{ab}+ E_{(2)}^{ab} \pa E_{(2)}^{ab}\nonu
 &&+m E_{(1)}^{ab} \pa E_{(1)}^{ab} ](w)
 + O(\frac{1
 }{(z-w)^2}),\nonu
 &&\Tt(z) E_{(1)}^{ab} E_{(2)}^{cd} E_{(2)}^{ac} E_{(2)}^{bd} (w)  =  \frac{
 1}{(z-w)^4} \frac{3}{(m+4)} [3(m-1) E_{(2)}^{ab} E_{(2)}^{ab}\nonu
 && +(3m+2) E_{(1)}
 ^{ab} E_{(2)}^{ab}](w)+\frac{
 1}{(z-w)^3} \frac{3(6m-1)}{2(m+4)} [ \pa E_{(1)}^{ab} E_{(2)}^{ab}\nonu
 &&- E_{(1)} \pa E_{(2)}^{ab}]
 +O (\frac{1}{(z-w)^2}),\nonu
 &&\Tt(z) E_{(2)}^{ab} E_{(2)}^{cd} E_{(2)}^{ac} E_{(2)}^{bd} (w)  =-\frac{
 1}{(z-w)^4} \frac{3}{(m+4)} [6(m-1) E_{(2)}^{ab} E_{(2)}^{ab}\nonu
 &&+(6m-1) E_{(
 1)}^{ab} E_{(2)}^{ab}](w)
 +\frac{
 1}{(z-w)^3} \frac{3(6m-1)}{(m+4)} [ E_{(1)}^{ab} \pa E_{(2)}^{ab}\nonu
 &&- \pa
 E_{(1)}^{ab} E_{(2)}^{ab} ](w)+O (
 \frac{1}{(z-w)^2}).
 \eea

 In general, we can not find spin 4 primary coset field consisting of 18
 fields under $\Tt(z)$ without $\Ut(z)$ because of too many unknown
 coefficients for 8 constraint equations. The results obtained so far
 can be summarized as follows: We have found 3
 currents of eq. (2), (5) and (10) in the coset model ( and in the $ c=5/2 $
 free
 fermion model ). It is cumbersome to proceed operator product algebras
 $ \Ut(z) \Vt(w) $ and $ \Vt (z)
\Vt(w) $ thoroughly.

 The Neveu-Schwarz sector of the 5 fermion model decomposes into two
irreducible
 highest weight modules of $\hat{so} (5)$ at level 1. In this circumstances one
obtains the following
 expression for the generating functional of states in the Neveu-Schwarz sector
 of the 5 fermion model that are singlets under $so (5)$ \cite{watts}.
 \bea
 \chi_{\mbox{singlets}}^{\mbox{NS}} (q) & = &
\frac{(1-q)(1-q)(1-q^2)(1-q^3)}{(1+q^{\frac{1}{2}})(1+q^{\frac{3}{2}})}
\prod_{k=1}^{\infty} \frac{(1+q^{k-\frac{1}{2}})}{(1-q^{k})^{2}}.
 \eea

The factor $(1-q)(1-q)(1-q^2)(1-q^3)$ in the numerator corresponds to
 the fact that the modes of $ L_{-1}, V_{-1}, V_{-2}$ and $ V_{-3} $
annihilate the vacuum
by requiring the regularity of the vacuum. In other words, they generate
singular vectors.
By counting the singlets in the factor module, vacuum irreducible Verma module,
we can
 recognize the generating currents that are in the chiral algebra
\cite{ba,watts}. Therefore we can easily read off the conformal dimensions.

\bea
 \chi_{\mbox{singlets}}^{\mbox{NS}} (q) & = & \frac{\phi_{\frac{5}{2}}
(q)}{\phi_{2} (q) \phi_{4} (q)}\nonu
					& = & 1+q^2+q^{\frac{5}{2}}+q^3+q^{\frac{7}{2}}+3q^4+2q^{\frac{9}{2}}\nonu
					&   & +3q^{5}+3q^{\frac{11}{2}}+7q^6+6q^{\frac{13}{2}}+
 8q^7+O (q^{\frac{15}{2}}),
 \eea
 where we have defined the ' modified Euler function '
 \be
 \phi_{\Delta} (q)=\left\{ \begin{array}{ll}
                   \strutje \prod_{k=\Delta}^{\infty} (1-q^k)
                   & \mbox{if $\Delta$ is integer}\\
                    \strutje \prod_{k=\Delta}^{\infty} (1+q^{k-1})
                   & \mbox{if $\Delta$ is half integer.}
			  \end{array}
		  \right.
 \ee
 All the states in the vacuum Verma module obtained by acting with the creation
 modes of the currents in the associated Casimir algebra are invariant
 under the
 finite horizontal subalgebra.
 Of course, this is backed up by the closed operator algebras \cite{ahn}.
 Eq. (17) gives the number of independent states at each level in the Fock
 space of 2 boson fields and a fermionic field.
 A basis for the 7 dimensional eigenspace of $so(5)$ singlets with $L_{0}$
 eigenvalue 6 is
 \bea
 L_{-6} | 0 >_{NS},\;\;\; V_{-6} | 0 >_{NS},\;\;\; L_{-2} L_{-2} L_{-2} | 0 >
 _{NS} \nonu
 L_{-4} L_{-2} | 0 >_{NS},\;\;\; V_{-4} L_{-2} | 0 >_{NS},\;\;\; L_{-3} L_{-3}
  | 0 >_{NS} \nonu
 \mbox{and}\;\; U_{-\frac{7}{2}} U_{-\frac{5}{2}} | 0 >_{NS}.
 \eea
 The explicit check of the absence of null state of dimension 6 is
straightforward.

 From the above argument of the character for an irreducible representation,
 the structure of Casimir algebra, $WB_{2}$ algebra,
 is generated by 3 currents,
 which have dimension $ 2, 4, 5/2 $ respectively.
The operator algebras of these currents are expected to be
 similar for the 5 free fermion model $and$ for the generic coset model.
 The coset models, at least for large $m$, can be viewed as perturbations of
the limit model at $ c=5/2$ and will share its algebraic structure.
This situation is similar to the case of \cite{ba}.
We propose the closure of the operator product algebra of Casimir operators
in the coset model $\hat{SO}(5)_{1} \times \hat{SO}(5)_{m} / \hat{SO}(5)_{1+m}
$ without the explicit calculations of the operator product algebras
$\Ut(z) \Vt(w)$ and $\Vt(z) \Vt(w)$. It would be interesting to relate the
operator contents of (2), (5) and (10) to those of free field construction
 \cite{fl2}.

 \vspace{15mm}

 This work was supported in part by grant NSF-91-08054.

 \end{document}